# GridLAB-D: An agent-based simulation framework for smart grids

David P. Chassin,[1,2] Jason C. Fuller,[1] and Ned Djilali[2,3]

[1] *Pacific Northwest National Laboratory, Richland, Washington, USA*
[2] *Department of Mechanical Engineering, and Institute for Integrated Energy Systems, University of Victoria, Victoria BC, Canada*
[3] *Faculty of Engineering, King Abdulaziz University, Jeddah, Saudi Arabia*

Correspondence should be addressed to David P. Chassin; david.chassin@pnnl.gov

Simulation of smart grid technologies requires a fundamentally new approach to integrated modeling of power systems, energy markets, building technologies and the plethora of other resources and assets that are becoming part of modern electricity production, delivery and consumption systems. As a result, the US Department of Energy's Office of Electricity commissioned the development of a new type of power system simulation tool called GridLAB-D™ that uses an agent-based approach to simulating smart grids. This paper presents the numerical methods and approach to time-series simulation used by GridLAB-D and reviews applications in power system studies, market design, building control system design, and integration of wind power in a smart grid.

## 1. Introduction

Recent smart grid technological advances present a new class of complex interdisciplinary modeling and simulation problems that are increasingly difficult to solve using traditional computational methods. Emerging electric power system operating paradigms such as demand response, energy storage, retail markets, electric vehicles and a new generation of distribution automation systems not only require very advanced power system modeling tools, but also require that these tools be integrated with building thermal and control models, battery storage technology models, vehicle charging system models, market simulators, and detailed power system control models. Historically all of these simulation tools were developed independently and each treated the others as a quasi-static boundary condition, an approach that not only limits their effectiveness in evaluating technology impacts over multiple scale and multiple time horizons, but which also neglects potentially important coupling effects.

In the case of electricity transmission simulation, PSLF [1], PSS/E [2] and Powerworld [3] have a longstanding record showing their ability to simulate bulk power systems in a wide range of conditions. But even with recent improvement to address new conditions such as fault-induced delayed voltage recovery [4], these tools are largely unable to integrate with the wide variety of tools needed to address distribution-level phenomena in a manner that meets the needs for smart-grid technology developers.

Electricity distribution level tools such as SynerGEE [5], WindMil [6], Cymdist [7], and RTDS [9] face similar challenges integrating with wholesale market and renewable integration tools because they too were designed using conventional models that depend on homogeneous descriptions of the underlying electromechanical behavior of the electric power system, either as an electromagnetic transient solution with timescale of microsecond to milliseconds, or as a steady-state power flow solution with no timescale at all. At intermediate timescales of seconds, minutes, hours and days there are many important phenomena that these simulations cannot incorporate.

The same can be said for building energy simulation, battery storage models, market simulations, and distribution automation controls in that they cannot represent the behaviors of the subject systems at all the time and size scales that each of the others require to work properly. Thus the problem of integrating these tools into a single multi-scale computational framework appears insurmountable, were we to restrict ourselves to the conventional simulations based on the numerical solution of systems of ordinary or partial differential equations (or their discretized counterparts) to represent changes in quantities of interest such as the voltage at a bus, the price of energy, the temperature in a building or the charge in a battery.

Agent-based computational economics [9] has increasingly been applied to electricity markets and consequently was among the first fields to address the challenges of using agent-based tools in power system simulation. The limitations with classical modeling methods [11], the tendency to ignore learning as a result of one-shot auctions [12] and the concerns with stylized trading models used by game theoretic methods [13] are among the chief motivations cited for using agent-based methods. The exploration of multiple equilibria [14] and a change from a focus on rational behavior and equilibrium processes toward heterogeneity and adaptation [15] only becomes possible with significant computing resources. As a result, computational economics has been divided into four areas of investigation [14]:

1. Empirical research that seeks to understand why and how macroscopic regularity emerge from microscopic properties and behaviors;
2. Normative research that uses agent-based models to as an *in silico* laboratory to design and test policies;
3. Theory generation that uses structured analysis to discover the conditions under which global regularity evolves; and
4. Methodology development that seeks to improve the tools and methods that support computational economics.

Consistent with the postulate that markets should be designed using engineering tools [16] and anticipating the coming smart grid revolution the US Department Energy's Office of Electricity commissioned Pacific Northwest National Laboratory to develop a simulation environment that would address the gaps in existing power system simulation and modeling tools. The first open-source release of the GridLAB-D [17] occurred in April 2008 and by November 2010 GridLAB-D was used to study a variety of smart grid problems in demand response and renewable integration [18]-[26]. Since then, the software has been improved and additional capabilities have enabled the study of a wide range of smart grid problems including conservation voltage reduction, microgrid control, retail market design, wholesale-retail market integration, distributed resource control, smart grid technology readiness evaluation, distributed energy resource integration, reduced-order model development, appliance control strategies, generation intermittency impacts on distribution systems, photovoltaic integration impacts, large-scale integration of wind power, smart grid cost-benefit analysis and transmission-distribution system model integration.

The primary purpose of this paper is to place the development of GridLAB-D in the context of research on the application of agent-based simulations and provide details of how GridLAB-D solves inter-disciplinary simulation problems as a time series using the agent-based paradigm. The first section discusses the general features of agent-based systems and briefly discusses examples and features of such systems in various domains. Next, the solution methods used by GridLAB-D are discussed and application examples are reviewed to demonstrate how the methods have been applied to various smart grid problems. Finally, planned future work is discussed and opportunities for other researchers to contribute further developments to the open source GridLAB-D tools are enumerated.

## 2. Fundamentals

Agent-based modeling is not a new approach to modeling complex systems. Early development of this approach was pioneered by Ulam and von Neuman [28], popularized by Holland [29] and Conway [30] and systematized by Wolfram [31]. But advances in computational capabilities in recent years have made large-scale agent-based models much more accessible to the non-experts using desktop computing systems. Agent-based simulations have become commonplace in games, finance, epidemiology, ecological research, and training systems to name a few examples. This section will examine a well-known example from ecology to elucidate the fundamental aspect of agent-based simulation.

To illustrate the difference between conventional models and agent-based models, we review the Lotka-Volterra predator-prey model [32], a well-studied class of system that has been modeled using both conventional and agent-based methods. A Lotka-Volterra system describes a simple ecosystem that exhibits quasi-harmonic behavior we can observe using simulations based on both methods and thus provides a good basis for comparison [33]. This well-known predator-prey system is described by the ordinary differential equations

$$\dot{x} = x(a - by) \quad (1a)$$
$$\dot{y} = y(cx - d) \quad (1b)$$

where $x$ is the size of the prey population at time $t$ and $y$ is the size of the predator population at the same time $t$. If $x$ represents the number of rabbit and $y$ represents the number of foxes, equation (1a) says that while rabbits grow at a rate $a$ they are also killed by foxes at a rate proportional to the number of foxes $b \cdot y$. Similarly, equation (1b) says that while foxes grow as a function of the food supply $c \cdot x$ they also die at a rate $d$.

The simplicity of the Lotka-Volterra system lends itself to analysis in the sense that one can compute aggregate properties such as the fixed population equilibrium by solving equations (1a) and (1b) for the steady state when $\dot{x} = \dot{y} = 0$. In this case we find only one non-trivial equilibrium state when

$$y = \frac{a}{b} \text{ and } = \frac{d}{c}. \tag{2}$$

Similarly, the stability of the fixed points can be determined using the Jacobian

$$J(x,y) = \begin{bmatrix} a - by & -bx \\ dy & cx - d \end{bmatrix} \tag{3}$$

from which we conclude the trivial fixed point $J(0,0)$ is an unstable saddle point, which explains why populations are not "attracted" to extinction conditions. The non-trivial fixed point is different because $J(d/c, a/b)$ has imaginary eigenvalues $\lambda_1 = i\sqrt{ad}$ and $\lambda_2 = -i\sqrt{ad}$ and no conclusions can be drawn. Solving the original differential equations by integrating directly allows us to find a conserved quantity

$$C = a \ln y(t) - b\, y(t) - c\, x(t) + d \ln x(t), \tag{4}$$

the value of which corresponds to a stationary population that oscillates around the non-trivial fixed point along invariant trajectories. Thus satisfying equation (4) provides the basis for any simulation that will accurately model the population dynamics based on the parameters $(a, b, c, d)$ and the initial conditions $x(0)$ and $y(0)$. Given this condition all possible states of the system can be explored, as shown in Figure 1.

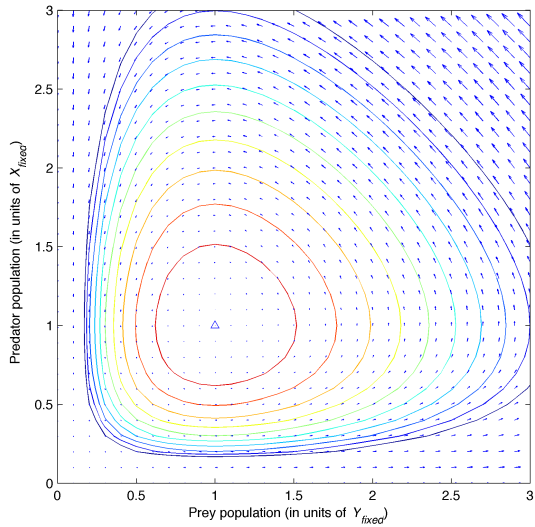

Figure 1: State space trajectories of a predator-prey system

Several significant problems become apparent when one attempts to find analytic solutions to many real-world systems. First, the parametric form of equation (4) is often difficult to solve numerically as a time-series solution, even for simple systems such as the Lotka-Volterra model, and finite difference methods often exhibit numerical integration errors that accumulate over time and lead to divergence, as shown in Figure 2. The source of this particular error is the estimate of the derivative at the start of each finite time interval. Euler's method addresses this problem to a first order and higher order solutions use Runge-Kutta methods to eliminate the error. Unfortunately, for many systems these error correction methods can be challenging to implement numerically using suitable finite difference methods.

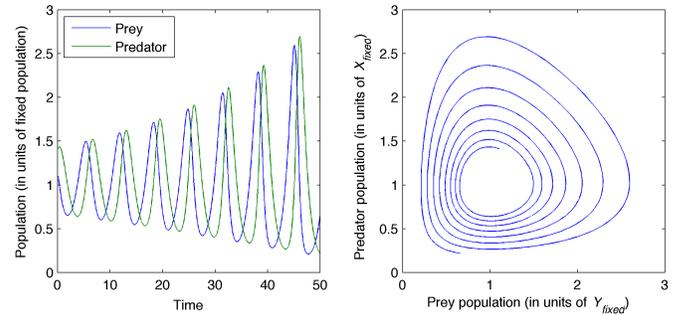

Figure 2: Integration error of a "naive" finite difference solution

The second problem is any change to the structure of the system or coupling with other dynamic systems requires that the solution be re-derived (often manually) from the original differential equations. This difficulty would be encountered were we to attempt to solve a mixed electro-mechanical, thermal and economic system such as

$$\tilde{I}_{N \times 1} = Y_{N \times N} \tilde{V}_{N \times 1} \tag{5a}$$

$$\tilde{V}_{L \times 1} \tilde{I}^*_{L \times 1} = Q_{L \times 1} R(P_{M \times 1}) \tag{5b}$$

$$\tilde{V}_{G \times 1} \tilde{I}^*_{G \times 1} = Q_{G \times 1} R(P_{M \times 1}) \tag{5c}$$

$$\frac{\partial D_{L \times M}}{\partial P_{L \times M}} = \frac{\partial S_{G \times M}}{\partial P_{G \times M}} \tag{5d}$$

where
- $\tilde{I}_{N \times 1}$ represents the phasor currents flowing into the network at the $N$ electric nodes,
- $Y_{N \times N}$ is the node admittance matrix,
- $\tilde{V}_{N \times 1}$ represents the $N$ node voltage phasors,
- $Q_{L \times 1}$ represents the $L$ customers thermal loads,
- $Q_{G \times 1}$ represents the $G$ generators outputs,
- $R(P_{M \times 1})$ represents the generator or customers responses to the $M$ prices for electricity (e.g., energy, power, ramping),

- $D_{L \times M}$ is the demand of $L$ customers in $M$ markets, and
- $S_{G \times M}$ is the supply of $G$ generators in $M$ markets.

The first equation describes the equilibrium condition for the electric power flow network, the second and third equations describe the equilibrium condition for the generator and consumer response to pricing and the fourth equation describes the equilibrium condition for the power market. These systems tend to operate on different timescales using different variables to describe interfaces between them.

Finally, the third problem is that as the systems become more complex the differential equations become so numerous and unwieldy that the model becomes analytically intractable for any non-trivial condition. This is certainly the case when power systems, market systems, and building thermal models are combined as above with equations (5a-5d).

Fortunately, no matter how complex these systems become they can be numerically modeled using agent-based methods. To illustrate how this is done we review a model of the same predator-prey systems using Conway's Game of Life, which uses cellular automata to represent a 2-dimensional landscape in which the populations interact. This landscape is represented by a matrix of cells that can take one of three values: 0 when a cell is vacant; 1 when it is occupied by a prey; and 2 when it is occupied by a predator. To simulate the advance of time the total population of each is counted as the matrix is iteratively updated using simple rules such as

1. The probability that a rabbit is born in a cell "adjacent" to a cell occupied by a rabbit is $p$.
2. The probability that a fox replaces a rabbit in a cell "adjacent" to a fox is $q$.
3. The probability that a fox dies is $r$.

where the meaning of "adjacent" uses a "north-east-south-west" cell adjacency condition that embodies the probability of the $xy$ component in the Lotka-Volterra model. But the 2-dimensional adjacency definition is unnecessary for the purposes of simulation. Adjacency can also be accomplished using a 1-dimensional "left-right" adjacency definition without loss of generality.

The output of an agent-based simulation of the Lotka-Volterra model based on these rules using a simpler random 1-dimensional "encounter" map generated at each iteration can be seen in Figure 3 for $p = 0.2$, $q = 1.0$, and $r = 0.2$. The fixed point for these conditions is $x = 10^4$ and $y = 5 \times 10^3$. The simulation produces similar oscillatory behavior to that observed in the analytic model. The model naturally introduces small fluctuations to which it is sensitive, which is why it deviates from the fixed point even when it is initialized at it. This phenomenon can be important in systems where the action of a single entity can influence the outcome of the entire system. Such situations are difficult to describe using ODEs because they involve fast-growing instabilities emanating from independent perturbations.

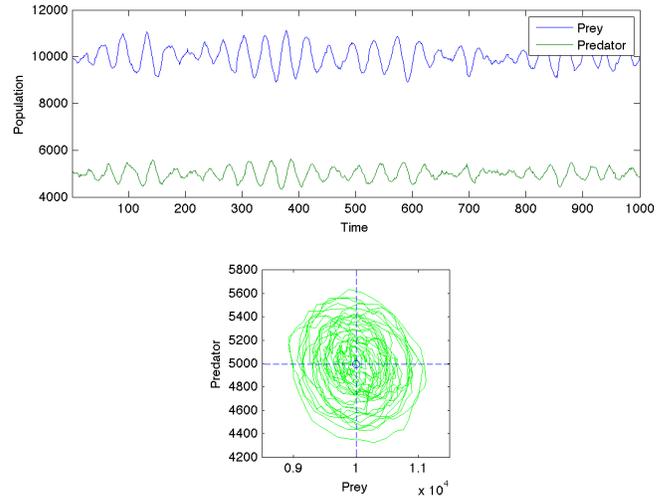

Figure 3: Fluctuation behavior of agent-based simulation

The simulation shown in Figure 3 exhibits another characteristic not seen in ODE solutions and which is shown in Figure 4 (only the initial conditions have changed). While finite difference solutions typically exhibit distinctly convergent or divergent behavior, agent-based solutions often exhibit mixed convergence behavior not seen in simpler ODE solutions.

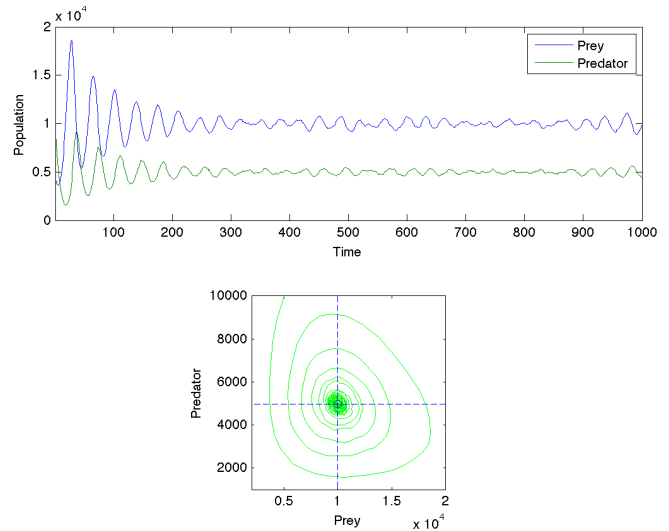

Figure 4: Convergent behavior of agent-based simulation

The primary characteristics of verisimilar agent-based models are based on realistically defining the agents and their relationships in an interaction landscape so that their evolution over time accurately reproduces the dynamics of the system being modeled. This requires that the following considerations be addressed carefully:

1. The internal states of the agents are represented by variables (discrete or continuous) that provide sufficient dynamic range and resolution to allow small fluctuations to affect agent behavior realistically;
2. The agents' behaviors are represented such that they evolve in a manner akin to a state machine (e.g., a Markov process, a cellular automaton, or a state space model) or equivalent model (e.g., differential equations)
3. The agents interact with an environment that evolves over time such that the agents are both affecting the environment and affected by the environment; and
4. The agents interact with each other in a manner that is consistent with the expected interactions in the system being modeled, i.e., not all the internal state variables of agent are revealed to other agents.

More complex systems may also involve the following additional considerations:

5. The environment may change over time, i.e., an external simulation, underlying model, or prerecorded set of conditions changes it slowly with respect to the dynamics of the agents.
6. Agents of different types can be interacting concurrently, which is the most common situation in highly realistic simulations.

Unfortunately, while the descriptive power of agent-based models is readily apparent and has been amply demonstrated [34]-[37], these models have a few significant shortcomings that remain for the most part unresolved. The first is that agent-based simulations provide even less analytic insight than numerical simulations that are derived on *ab initio* models. For example, the simulation shown in Figure 3 and Figure 4 has an obvious fixed point that corresponds to what we expect from the Lotka-Volterra model. However the simulation does not provide us with an analytic relationship between the fixed point we observe, which is easily found numerically by taking the mean values of $x$ and $y$ over a non-trivial range of time and the parameters of the analytic model or the probabilities of the rules. While both $x$ and $y$ can be determined analytically from the Lotka-Volterra parameters using equation (2), there is no obvious way to relate the fixed point to probabilities $p, q$, and $r$ without running the agent-based model. This challenge remains unresolved except for the most trivial system.

The second shortcoming is that agent-based model verification and validation is difficult to accomplish using formal methods. When considering conventional simulations such as the ODE solution to the Lotka-Volterra system, verification is the process of ensuring for example that the populations always satisfy equation (4) given any particular initial condition $x(0)$ and $y(0)$, whereas validation is the process of ensuring that a series of observations of a real population evolves in a manner consistent with predictions of the simulation.

In principle the same should be possible with agent-based simulations. However, as we have already seen agent-based model often exhibit fluctuations that resemble real systems so that instead of trying to verify an idealized model (non-fluctuating) against real (fluctuating) data, model developers are often trying to validate using two systems that may fluctuate in different ways. The verification question is no longer just about the uncertainty associated with the empirical data. Now the uncertainty associated with the agent-based model must be considered as well. Even for simple models like the Lotka-Voltera system the agent-based model does not strictly obey its conservation law: it only approximately follows the law in the sense that the simulation converges toward the fixed point when far from it but diverges when very close to it.

The problems with agent-based model validation came to the fore in the development of agent-based computational economics used in the design of electricity markets. In their discussion of this problem LeBaron and Tesfatsion identified three challenges [38]. Foremost is the embarrassing number of degrees of freedom that arise from the large number of parameters contained in the models (a simple 1000-home GridLAB-D model contains nearly 200,000 distinct parameters that can affect the outcome). This problem is made all the more severe by the nearly unlimited functional and learning algorithms that can be implemented, and even mixed. Finally the properties of the agents themselves are often difficult to identify precisely and are often informed more by modelers intuition and engineering expertize than by data collected or observed human behavior.

One commonly used approach is verification against experimental and/or reduced simulations and validation against empirically collected data that demonstrate consistency with known initial conditions and outcomes. In this way, GridLAB-D models are often verified with simple "known good" simulations and validated using telemetry from large-scale real-world systems for which models are available. The latter is discussed further in the applications section below.

These problems are not unique to GridLAB-D. Widely-used agent-based simulation environments such as SWARM [34], Repast [35], EMCAS [36], AMES [37] and others have all addressed these challenges and the reader is referred to these for details on how the verification and validation problem is addressed variously by them.

Ultimately the decision whether to accept the mathematics of agent-based models hinges on an argument made by Borrill and Tesfatsion in their assessment of the relevant differences between classical and constructive mathematics [38]. The former is supported by those who accept the law of excluded middle so that existence proofs by contradiction as permissible. While the latter is supported by those who require direct proof that a proposition is true in order to rule out both falseness *and* undecidability. This latter proof can be realized as computer programs that embody concepts of information flow and limits on what is known when by any agent:

*"This distinction provides a dramatically different perspective on how we perceive models in our mind in relation to the real-world systems they are intended to represent. For example, social system modelers using classical mathematics typically assume (explicitly or implicitly) that all modeled decision makers share common knowledge about an objective reality, even if there is no constructive way in which these decision makers could attain this common knowledge. In contrast, social system modelers advocating a constructive mathematics approach have argued that the "reality" of each modeled decision maker ought to be limited to whatever that decision maker is able to compute."*

GridLAB-D was designed and implemented with the latter view in mind.

## 3. Solution method

The success of GridLAB-D as a tool to study smart grids is primarily attributed to the use of the agent-based simulation paradigm. The approach has made GridLAB-D easy to use in spite of the extensive use of multi-disciplinary elements in various modules. In addition the output of GridLAB-D is highly similar to data collected from smart grid demonstration project conducted in the field, which has facilitated verification and validation.

GridLAB-D allows modelers to choose which of the agent-based characteristics are implemented in a given module. Multiple modules may be operated concurrently in any given simulation and there is no requirement that every module use the same modeling method in any given simulation. For example, the powerflow module uses a state-space model with an underlying algebraic solver to compute the voltages and currents given the loading conditions presented to it by the other modules. The residential building module in turn use the voltages to determine the energy input to home energy systems and an underlying ODE thermal model to solve the indoor air and mass temperatures. The market models use the building control systems to determine the price of electricity, which in turn is used to determine the price at which overall supply and demand for electricity are equal.

Agents are organized into ranks based on the relationships between them imposed on them by the modules' solvers. The ranks are organized in trees of parent-child relationships, with each parent agent primarily depending on the values accumulated from one or more child agents. The ranks are given ordinal numbers with the greatest ordinal assigned to the agent that is the topmost rank and zero assigned to the bottom-most rank. In practice it is typical to find only one agent at the topmost rank and a plurality of agents of rank zero.

The determination of the ranks is made by the modules and based on the solution method implemented. For example, the *powerflow* module uses a different rank structure depending on whether the forward-backsweep method [40] and current injection [41] method is used. When the forward-backsweep method is used, the rank structure tends to require many ranks whereas when the current injection method is used only two ranks are required. This difference is known to influence the relative performance of the solvers depending on the size and general structure of the electric network being modeled.

To illustrate how GridLAB-D's solver gathers values from multiple agents consider how the average

$$y = \frac{1}{N}\sum_{n=1}^{N} x_n \quad (6)$$

is computed. There are three steps required to complete this operation:
1. set $y = 0$ and $N = 0$
2. add each $x_n$ to $y$ for $n = 1, 2, \ldots, N$ incrementing $N$ each time
3. divide $y$ by $N$ if $N$ is non-zero

This process can be conducted for $N$ values of $x$ in three phases, within which multiple operations (if any) can be conducted in parallel provided the operations on $y$ and $N$ are atomic, as shown in Figure 5.

$$\underline{Phase\ 1} \quad \underline{Phase\ 2} \quad \underline{Phase\ 3}$$

$$\begin{array}{ll} & y \leftarrow y + x_1, n \leftarrow n + 1 \\ y \leftarrow 0 & y \leftarrow y + x_2, n \leftarrow n + 1 \\ n \leftarrow 0 & \vdots \\ & y \leftarrow y + x_N, n \leftarrow n + 1 \end{array} \quad y \leftarrow \begin{cases} indet. & : n = 0 \\ \dfrac{y}{n} & : n > 0 \end{cases}$$

Figure 5: Three parallelized phases to compute $y = \frac{1}{N}\sum_{n=1}^{N} x_n$.

In GridLAB-D parlance, these phases are denoted
1. **pre top-down pass**: this phase gives agents the opportunity to prepare to receive updates from other agents;
2. **bottom-up pass**: this phase gives agents the opportunity to update other agents; and
3. **post top-down pass**: this phase gives agents the opportunity to compute final values based on updates received from other agents.

In addition, GridLAB-D includes modules that allow implementation of so-called "precommit" and "commit" passes before and after the three main phases, respectively. This permits solvers that collect global data to prepare and commit the global data that are affected, if any. There is also a finalize pass that complete only when the clock is advanced to allow any objects that need to compute time-dependent updates to do so before the clock is advanced.

A number of built-in properties with special characteristics are provided to represent physical or stochastic processes and maintain endogenous or exogenous relationships. These are always updated before the pre-commit stage to ensure that all the global, module and object properties are correct at a given time indicated by the global clock. These built-in properties are updated in the following order

- links to external simulations (both read and write)
- random variables (based on the supported distributions provided by GridLAB-D)
- scheduled values (updates on the ISO "cron" schedule standard)
- loadshapes (primarily used to shape values in time using queues, pulse-width modulation, amplitude modulation or simple analog shapes)
- transforms (functions that update a property based on the values of other properties)
- enduses (structures that describe how electric loads are composed)
- heartbeats (events that occur on a regular basis independent of synchronization events)

All these updates (with the exception of finalize updates) return a time for the next expected event for the object or property in question.

### 3.1. Convergence and Synchronization

The current implementation of GridLAB-D does not guarantee convergence in the overall solution. Modelers can create situations in which two or more agents cannot find a combination of states that satisfy their respective convergence criteria. The design of GridLAB-D's overall solver assumes that such situations are not intentional but due to modeling error. In this case GridLAB-D detects the resulting large iteration count without the global clock changing and stops the simulation. Modelers must implement non-iterative methods of resolving such state conflicts based on the assumption that they occur at a time scale less than the time resolution of the main solver.

If necessary modelers can process events with a time resolution less than 1 second using a discrete fixed timestep solution method. The discrete timestep used is the shortest timestep requested by the agent(s) seeking subsecond processing. During processing of subsecond simulations event-driven simulation is disabled until all agents indicate that they no longer require subsecond processing, which typically occurs when the transient behavior settles to steady state and event-based processing can resume.

Parallelization of many event and timestep update computations is accomplished using a "thread group" strategy. This multiprocessor approach to improving simulation performance preallocates computational threads to groups of object event handlers that the model loader determines can always be executed independently. The determination of parallelizability is based on the rank of an object and the type of event being processed. Ranks are established on the basis of which way information flows during synchronization events, if any, with high-rank objects depending on multiple low-rank objects during the bottom-up synchronization event. This method of parallelization has been shown to exhibit approximately linear scaling for the smaller number of computing units typically found in desktop computing systems [42].

Synchronization avoidance strategies are also included in GridLAB-D's main solver to reduce the number of unnecessary events and improve overall simulation performance. Among these is the "valid to" time, which agents may set when the time of the next event in an agent is independent of external inputs to the agent and the agent is not expecting to make any changes to its internal state until that time. Using this valid-to time the main solver can avoid processing certain agent event when the outcome of the call is a forgone conclusion and the agent is not expected to change state.

### 3.2. Standard Modules and Solution Methods

The AC powerflow solution is implemented using a modified Newton-Raphson method for meshed electric networks [43]. The powerflow solver support unbalanced three-phase networks. The modeler does not need to directly compute the admittance matrix as the solver performs this update during synchronization from the properties of the objects that represent the various electrical supported by GridLAB-D powerflow module, including power lines, transformers, switchers, capacitor banks, voltage regulators, etc. In cases where the electric system is radial the modeler may opt to use the powerflow module's forward backsweep solver, which is based on Kersting's method [44]. A separate generators module provides classes that allow modelers to implement various electricity generating and storage resources.

Building thermal response is solved using the equivalent thermal parameters (ETP) method, which implements both the time and temperature solutions to the second-order ordinary differential equation that describes the response of the indoor air temperature of a building to outdoor temperature conditions, internal appliance and occupant heat gains, ventilation gains/losses, solar gains and heating/cooling system state. The building modules include thermostatic controllers and appliance models some of which incorporate demand response control strategies. A retail electricity market is included that implements a double-auction for feeder capacity and determines the real-time price (RTP) at which feeder supply is equal to the total load. The retail market support both demand response resources, such as thermostats and electric vehicle chargers, as well as distributed generation resources such as diesel backup generators, microturbines, photovoltaics and energy storage devices.

## 4. Applications

Grid-LAB-D has been used to study a wide variety of power system problem as summarized above. In this section we examine a few of the results obtained in more detail and discuss the role that GridLAB-D's solution method played in enhancing the analysis beyond what is possible using conventional simulation tools. Specific applications or recent interests include Volt-VAR Optimization (VVO), dynamic real-time pricing (RTP) experiments, and integration of renewable energy aided by demand response. In some cases, reliable solutions can be found from other methods, while in others, the agent-based methodology provides unique insight. As validation is often a key question in agent-based systems, studies that have been validated against experimental field data will be discussed.

*4.1. Volt-VAR Optimization*

VVO is a traditional utility technique for reducing energy consumption or peak demand on an electric circuit by lowering the system voltage to the lower portion of the operational voltage band [45]. With the proliferation of communication and sensor equipment, the control and optimization techniques have become increasingly more complex, but there are always questions about the trade-off between the cost of a more complicated system and the benefits. GridLAB-D was used to estimate these benefits prior to deployment [46], and simulations crossed the boundary between power system and load behavior.

Each of the loads within the system is modeled as a complex process driven by inputs such as outside air temperature, occupancy, thermostat set points, etc. Each load is modeled as an agent with its own specifications and inputs. As the voltage varies throughout the distribution system, this is also used an input into the load model to quantify how a change in system voltage affects the energy consumption of the individual devices, and thus the overall system. For example, an electric water heater is essentially a resistive element at any given moment in time. As voltage is reduced, the power demand decreases. However, because of the closed-loop thermostatic control on the device, the same amount of energy is required to heat the water. This affects the duty-cycle behavior of the water heater, and is tracked through time by each agent. The effect is that peak load (i.e., the maximum number of devices in the *on* state simultaneously) is reduced, but energy consumption (i.e., the cumulative time the loads are in the *on* state) is not. Each model, whether an air conditioner, dishwasher, or other appliance, has its own inputs and responses to changes in voltage. Traditional models, which are basically linear, time-invariant solvers, are not able to easily capture these effects.

In a study with American Electric Power (AEP), new VVO technology was tested in GridLAB-D on eight distribution circuits. The technology was then deployed and tested on those same circuits. Simulation predicted a 2.9% average reduction while deployment produced a 3.3% average reduction in energy consumption. However, there were significant differences on individual feeders that could be attributed to changes in load composition between the modeling and testing phases.

*4.2. Real-Time Pricing Demonstration*

As part of the American Recovery and Reinvestment Act, a real-time pricing experiment was devised to engage consumer loads with five-minute energy prices, in-home displays, and equipment utilizing automated response [47]. The automated system requires individual air conditioners to construct a market bid that reflects their desire to run during the next market period (every five minutes). A central auction collects all of the bids, clears a market price, and then broadcasts a (single) price signal to all of the devices. In turn, the devices respond to this signal in a coordinated (but non-communicative) manner by modifying the behavior of the thermostat. The end goal is to reduce overall energy costs, both for the customer and the utility, and to reduce demand when there is a physical constraint on the system. GridLAB-D was used to develop and fine-tune the control and communication requirements prior to deploying this system [48].

The power system is modeled along with individual appliance loads. In addition, each thermostat is modeled with additional market controls acting as bidding agents, and a centralized auction agent collects all of the information and dispatches a price signal. When required, the communication system is also modeled separately (i.e., message packets are dispatched through a communication layer rather than directly within the software), including message delays and dropped packets [49]; each of the agents is responsible for understanding what to do when information is lost or delayed.

By working through the design of the control system (both the distributed bidding agents and the centralized auction) via the simulator prior to deployment, a number of affects and impacts could be evaluated and re-designed. Affects that caused undue strain on the system (such as synchronization of loads in response to the price signal, errors in load prediction, or loss of data) and potential impacts (such as changes in customer bills, equitable rebate and incentive mechanisms, or violation of local constraints) were evaluated and modified. For example, it was found through simulation that a slight error in the agent bidding caused by the thermostat deadband could cause system oscillations in power demand under certain circumstances. This was potentially a major flaw in the control system, and was corrected prior to construction of the thermostats that were deployed. Reports comparing field demonstration results to GridLAB-D simulations will be available in late 2014. The combination of linear and non-linear solutions, non-continuous state variables, binary operations, and sorting algorithms are not solvable through direct solution methodologies.

### 4.3. Demand Response for Renewable Integration

Environmental concerns have spurred a significant growth of electricity generation from wind power and other renewable energy sources in the last decade. The temporal and spatial variability of these resources present a number of challenges to power system operators, particularly with respect to power system reliability and reserve requirements. These techno-economic challenges have typically limited wind penetration to at most 30% of grid generation. Smart grid technology now makes it possible to address these challenges using novel strategies such as demand response, whereby loads can be controlled in response to power imbalance or market price signals, and adjust their power demand. This essentially shifts part of the burden of balancing power from the supply to the demand side and results in a reduction of costly contingency reserves.

Using GridLAB-D in conjunction with MATLAB, Williams presented a smart grid model to assess the potential of mitigating fluctuations associated with distributed wind power by using self-regulating, thermostatically controlled heat pumps. The modeling framework for the smart self-regulating system is shown in Figure 5.

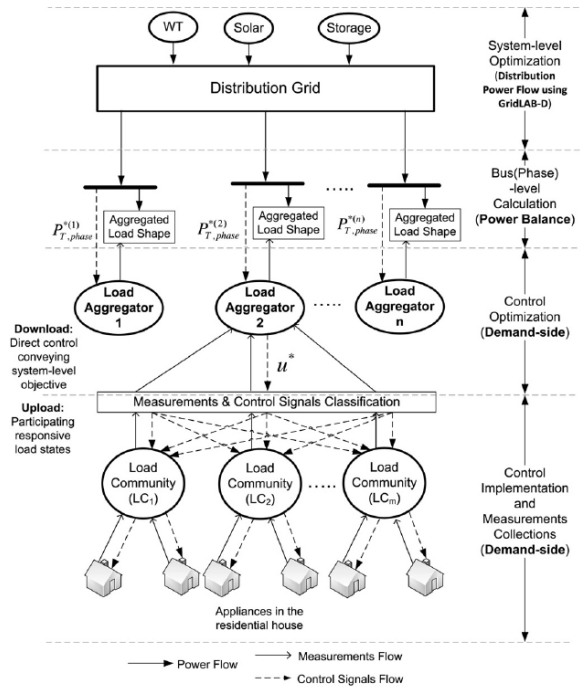

Figure 5: GRIDLab-MATLAB Modelling Framework for a Smart Self Regulating System (reproduced from [63] with permission from Elsevier)

Different bus-level control algorithms were investigated using the model, and Figure 6(a) illustrates the effectiveness of bus-level distributed heat pump management strategy in reducing load flow fluctuations by adjusting the heat pump demand to follow wind generation. A useful way of measuring the level of mitigation of wind fluctuations is to examine the required ramping rates spectrum of probability distribution [51]. Figure 6(b) shows the significantly lower ramp rates achieved by the control strategy.

The viability of demand response will require effective market pricing and the integration of price signals and load controls. Brooer et al [52] developed a general simulation framework integrating a GridLAB-D smart grid system with a market model. The model incorporates generator and load controllers, and allows bidding from both the supply and demand side into a double-auction RTP electricity market. Demand response in the system is achieved through thermostatically controlled loads such as Heating, Ventilation, and Air-Conditioning (HVAC) units, and electric water heaters.

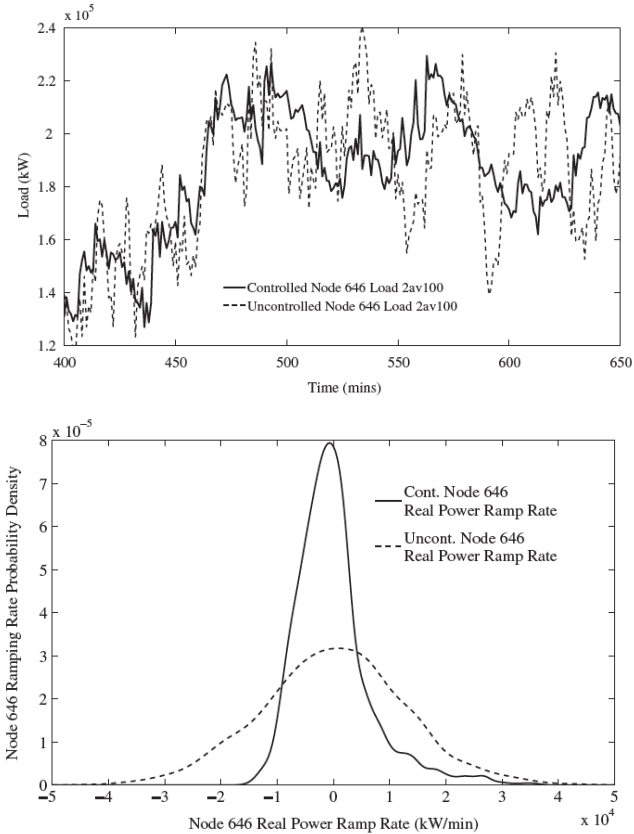

Figure 6: Effect of control on (top) load, and (bottom) real power ramp rate (reproduced from [50] with permission from Elsevier)

This model was validated using a physical demonstration project conducted on the Olympic Peninsula, Washington, USA. RTP Simulation results obtained using a system comprising 10,000 residential houses and a grid-integrated 35MW wind park are shown in Figure 7. The wind and hydro supplies consistently bid at $0/kWh and $0.1/kWh. respectively. The figure shows the response of a sample residential house to wind power variations. Bidding on the demand side is from the responsive HVAC load. The HVAC load switches off when the bid is below the clearing price. This happens when the clearing price rises as a result of decreasing wind power. The HVAC system switches on again when wind power recovers. This type of model allows operators to assess the impact of extreme scenarios, such as the persistence of high/low wind regimes over extended periods, and during which a diversity of loads needs to be maintained in order to avoid saturation (all loads becoming unresponsive).

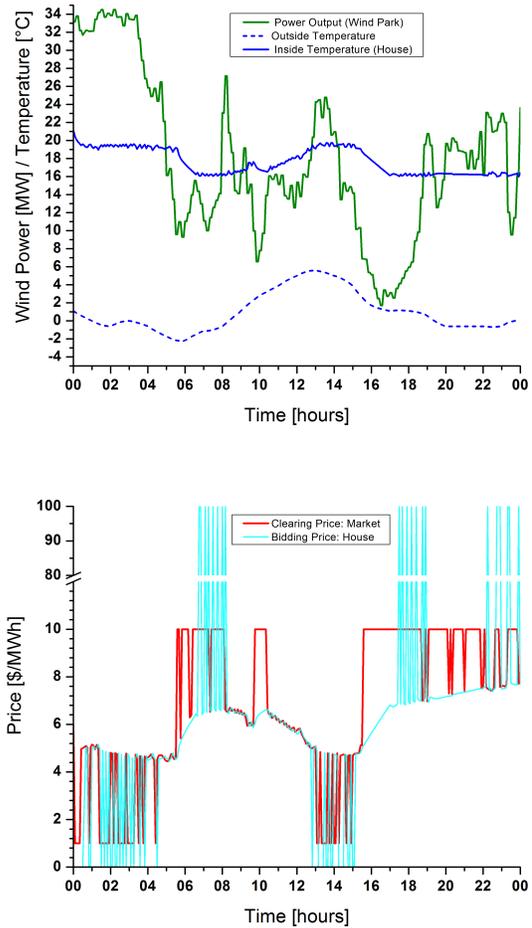

Figure 7: Controlled behaviour of an individual house over a 24h period in response to varying wind power. Top: Indoor house temperature following wind power. Bottom: Variation of market clearing price and resulting turning off of loads as a result of changes in wind power (adapted from Broeer et al. [52]).

These GridLAB-D-based renewable energy integration studies indicate that controlled customer power consumption can be modified to facilitate wind energy integration without compromising customers comfort. Such DR strategies can effectively modify load flow, improve energy efficiency and reduce contingency reserve requirements. The versatile GridLAB-D agent-based modelling provides an integrated framework to assess the potential of various demand response strategies and to support the design of Virtual Power Plants that can effectively provide the additional contingency reserve and regulation capacity required to increase the penetration of variable renewable energy generation in the electricity grid. Again, this cannot be accomplished through the use of traditional solution methods without making gross over-simplifications of the underlying behavior and response systems.

# 5. Future work

As an open-source tool for the smart-grid research community there are many prospective contributors to GridLAB-D development and thus many directions which it can go. The US Department of Energy's Office of Electricity, which currently directs the development of GridLAB-D at Pacific Northwest National Laboratory (PNNL) is committed to ongoing improvements in the power system, building, markets, controls and telecommunications modules themselves. In addition, PNNL is making investments in internal solution methods, with special consideration being given to parallelization for high-performance computing platforms, co-simulation environments to simplify integration with multiple simulation environments , and improvements to allow more formal model verification and validation methods.

# 6. Conclusions

In this paper we have presented previously unpublished details on the agent-based simulation methods implemented in GridLAB-D. Our objective is in part to introduce the applied mathematics community to the challenges faced by those who employ agent-based methods and encourage greater collaboration between applied mathematics and power engineering communities. We have presented the rationale for adopting an agent-based simulation approach to simulating the smart grid, discussed some of the challenges with using agent-based methods to design inter-disciplinary smart-grid technology solutions and reviewed at a high-level some applications and studies that best exemplify the versatility and impact of GridLAB-D.

# Acknowledgments

This work was funded in part by the US Department of Energy's Office of Electricity.